\documentclass[10pt,journal,compsoc]{IEEEtran}

\usepackage{amsmath,amsfonts}
\usepackage{algorithmic}
\usepackage{algorithm}
\usepackage{array}
\usepackage[caption=false,font=normalsize,labelfont=sf,textfont=sf]{subfig}
\usepackage{textcomp}
\usepackage{stfloats}
\usepackage{url}
\usepackage[hidelinks]{hyperref}

\usepackage{verbatim}
\usepackage{graphicx}
\usepackage{cite}
\usepackage{cleveref}
\usepackage{adjustbox}
\usepackage{multirow}
\usepackage{xcolor}

\begin{document}

\title{CarbonTag: A Browser-Based Method for Approximating Energy Consumption of Online Ads}

\author{José González-Cabañas\textsuperscript{†},
        Patricia Callejo\textsuperscript{‡,†},
        Rubén Cuevas\textsuperscript{‡,†},
        Steffen Svartberg\textsuperscript{§},\protect\\
        Tommy Torjesen\textsuperscript{§},
        Ángel Cuevas\textsuperscript{‡,†},
        Antonio Pastor\textsuperscript{§},
        and~Mikko Kotila\textsuperscript{§}% <-this % stops a space
\thanks{
\textsuperscript{†}UC3M-Santander Big Data Institute, Universidad Carlos III Madrid, Spain. \\ %, Getafe 28903, Spain. \protect\\
\textsuperscript{‡}Department Telematic Engineering, Universidad Carlos III Madrid, Spain. \\%, Leganés 28911, Spain.\protect\\
\textsuperscript{§}Cavai, 0152 Oslo, Norway.\protect\\}% <-this % stops an unwanted space
\thanks{© © 2023 IEEE. Personal use of this material is permitted. Permission from IEEE must be obtained for all other uses, in any current or future media, including reprinting/republishing this material for advertising or promotional purposes, creating new collective works, for resale or redistribution to servers or lists, or reuse of any copyrighted component of this work in other works.}

\thanks{\protect\\ Journal Reference: J. González-Cabañas et al., "CarbonTag: A Browser-Based Method for Approximating Energy Consumption of Online Ads," in IEEE Transactions on Sustainable Computing, 2023. DOI: 10.1109/TSUSC.2023.3286916.}}

%\markboth{IEEE Transactions on Sustainable Computing,~Vol.~X, No.~X, Month~Year}%
%{J. González-Cabañas, P. Callejo \MakeLowercase{\textit{et al.}}: CarbonTag: A browser-based method for approximating energy consumption of online ads}

%\IEEEpubid{0000--0000/00\$00.00~\copyright~2022 IEEE}

\newcommand{\carbonTag}{CarbonTag }
\newcommand{\adtag}{ad tag }
\newcommand{\adtech}{ad tech }

\newcommand{\new}[1]{{\color{black}#1}}

\maketitle

\begin{abstract}
Energy is today the most critical environmental challenge. The amount of carbon emissions contributing to climate change is significantly influenced by both the production and consumption of energy. Measuring and reducing the energy consumption of services is a crucial step toward reducing adverse environmental effects caused by carbon emissions. Millions of websites rely on online advertisements to generate revenue, with most websites earning most or all of their revenues from ads. As a result, hundreds of billions of online ads are delivered daily to internet users to be rendered in their browsers. Both the delivery and rendering of each ad consume energy. This study investigates how much energy online ads use \new{in the rendering process} and offers a way for predicting it as part of rendering the ad. To the best of the authors' knowledge, this is the first study to calculate the energy usage of single advertisements \new{in the rendering process}. Our research further introduces different levels of consumption by which online ads can be classified based on energy efficiency. This classification will allow advertisers to add energy efficiency metrics and optimize campaigns towards consuming less possible.
\end{abstract}

\begin{IEEEkeywords}
Ad, online advertising, energy consumption, carbon emissions, sustainability.
\end{IEEEkeywords}

\section{Introduction}
\label{sec:introduction}

\IEEEPARstart{T}{he} vast amount of energy consumed by human society is one of the biggest challenges faced by humanity today and in the near future. The production and consumption of energy contribute significantly to the emissions of greenhouse gases, including carbon, which contributes to climate change. The Statistical Review of World Energy 2022 by BP~\cite{bp2022} shows that the primary energy demand increased by 5.8\% in 2021, topping 2019 levels by 1.3\%. This indicates that the global need for energy is continuing to rise. The International Energy Outlook 2021~\cite{eia2022} predicts that over the next 30 years, the world's energy demand will rise by roughly 50\%. 

While there is a growing interest in using clean renewable energy (the amount of renewable energy consumption growing by over 2000 TWh between 2019 and 2021), polluting fossil fuels consumption remained largely unaltered \cite{bp2022} and they are expected to be used beyond the decade 2050 \cite{eia2022}, at which point the renewable energy is expected to become the most representative source of energy \cite{hubler2013eu}.

The transition from fossil fuels to clean energy is progressing more slowly than is necessary. It is paramount to implement strategies focused on reducing energy consumption to limit and prevent adverse environmental, social, and economic impacts, such as those associated with rising temperatures, rising sea levels, and decreasing precipitation. In addition, these strategies have recently become an immediate necessity in Europe, and the whole world, due to the energy crisis aggravated by the Ukraine War in 2022.

Since the advent of the internet in the early 1990s, the Information and Communication Technologies (ICT) sector has become a major driver of developed economies as well as a factor contributing to the increase in global energy consumption. Depending on the source, the total energy consumption of the Internet is approximated to be between 5\% and 15\% of total global energy consumption, that's is around 416.2 TWh per year~\cite{aslan2018electricity, websitecarbon, conversation:internet_consumption, nature:electricity}. This percentage is projected to rise sharply in the coming years.

Online advertising is one of the main sources of revenue supporting the operation of the commercial Internet. Millions of websites, social media platforms, mobile apps, video platforms, etc, rely on online advertisements as their main source of revenue. As a result, each day, hundreds of billions of online ads are delivered to Internet users to be rendered by their devices. Indeed, online advertising is a \$450 billion per year industry with an inter-annual growth of 35.4\% between 2020 and 2021, according to the \new{Interactive Advertising Bureau} (IAB)'s most recent Internet Advertising Revenue Report of 2021~\cite{IAB2021}. This marks the industry's highest growth since 2006 and highlights the importance of online ads as a fundamental driver of the Internet-age economy.

The online advertising industry has developed a complex technology ecosystem, known as \emph{\adtech} ecosystem, which is able to deliver personalized ads in nearly real time. This process may involve, for each ad delivery transaction, up to hundreds of different players with their associated data communications and processing events.

Previous literature has used high-level approaches to estimate that online advertising is a relevant contributor to the overall energy consumption of the Internet \cite{parssinen2018environmental, simons2010hidden, prochkova2012energy}. Therefore, without a doubt, online advertising must contribute to reducing the Internet's energy consumption by adopting more energy-efficient and therefore sustainable practices. The first step towards this consists in \emph{defining accurate measurement techniques able to measure the energy consumption of the ad delivery process, if possible, at the granularity of individual ads.}

The complexity of this goal, requires, in order to succeed, to employ a \emph{divide and conquer} approach. To this end, we have divided the ad delivery process into two separate parts: the \emph{network} and the \emph{device}. The \emph{network} part involves all the data processing and communications among intermediaries participating in the delivery of ads, whereas the \emph{device} part is defined by the ad rendering process executed by software installed in the device such as web browsers or mobile apps.

In this paper, we focus on the \emph{device}, thus \emph{our goal is to define an accurate measurement technique able to estimate the energy consumption of the rendering process of individual ads}.\footnote{Note that in the remainder of the paper when we refer to ad energy consumption we are implicitly referring to the ad rendering process occurring in the device.} We left the \emph{network} part for future research. \new{We conjecture the \emph{network} part may be responsible for an importantly larger fraction of the overall energy consumption of the ad delivery process in comparison to the ad rendering process}. However, properly measuring the \emph{device} part is relevant since (as we will see) the aggregated energy consumption of the ad rendering process across hundreds of billions of ads each day is not negligible. Moreover, it will serve to create awareness among both advertisers and Internet users, which are the fundamental players fueling the online advertising ecosystem. 

The goal of our paper is to define a methodology that allows measuring the energy consumption of the ad rendering process in real time and at scale. Ideally, one would wish to define a technique that measures the exact energy consumed resulting from executing the ad rendering process. Unfortunately, directly measuring the energy consumed by software processes such as ad rendering is not possible in practice (See Section \ref{sec:background}). 
Instead, the alternative is to define techniques to produce accurate estimations of the energy consumption of the process. To this end in this paper, we present the two following contributions that allow providing an accurate estimation of the energy consumption of the ad rendering process in real time and at scale.

\new{First, we have run controlled lab experiments to measure the energy consumption of over 25k ads using different hardware and operating systems. The results of this experiment were used to create a Machine Learning (ML) model that estimates the energy consumption of individual ads based on measurable parameters during the rendering process. The model's performance was found to be good, with R$^2$ values ranging from 0,67 to 0,97.}

\new{Second, we have built a system called \carbonTag, which uses the ML model to estimate the energy consumption of ad rendering in real-time and at scale. The system consists of an \adtag that collects the ad's parameters and sends them to the backend server. We have conducted lab experiments to evaluate the system's scalability and availability for integration by advertisers and \adtech intermediaries. Finally, we also propose an energy labeling system inspired by the EU energy labeling framework.}
%\end{itemize}
\section{Background}
\label{sec:background}

\subsection{Online Ad Delivery and Rendering Process}
\label{subsec:online_advertising}

The current online advertising ecosystem implements what is referred to as \emph{programmatic advertising}. When a user reaches a webpage (or mobile app), each ad space available on such webpage initiates the ad delivery process. The ad space sends an ad request message to the publisher's ad server requesting an ad to be placed in the ad space. The publisher's ad server can respond with a URL from where to receive the ad if a direct deal with an advertiser exists. Otherwise, the ad server forwards the ad request to an intermediary, which can be either an \emph{ad network} or a \emph{Supply Side Platform (SSP)}. The publisher's website and ad server, along with the SSP or ad network\footnote{Note that several ad networks or SSPs can be involved in the ad delivery process.} form what is referred to as the \emph{Sell Side} in programmatic advertising, since they are the entities involved in selling ad spaces. The term SSP is oftentimes interchangeable with the term Ad Exchange (AdX).
Using the information included in the ad request message \emph{bid request}, such sell-side actors process messages as described in the OpenRTB standard~\cite{openrtb}. The \emph{bid request} message is then sent to several \emph{Demand Side Platforms (DSPs)} initiating a real time auction process. DSPs are technology platforms where advertisers (or their agencies) create, manage, and deliver ad campaigns. A DSP would check the information included in the received bid request, e.g., location of the device, profile of the user (age, gender, interests), type of ad space (banner, video, etc.), venue of the ad space (web domain or mobile app), etc. If these parameters match the criteria defined in any ad campaigns on the DSP, the DSP responds with a \emph{bid response} including a bidding price for the ad space. The AdX runs a separate auction among the received bid responses and chooses a winner. The winner provides AdX with the URL of the ad to be delivered, which is hosted in the advertiser or DSP's ad server. This URL is forwarded back through the AdX, SSP/ad network, and publisher's ad server to the browser (or mobile app), which will render the ad. 

The described process defines the \emph{network} part of the ad delivery process. In addition, there are other subsidiary processes such as the tracking\&profiling process. The \adtech has developed a sophisticated ecosystem to track users' activity online (but also offline, for example tracking users' locations), enabling the profiling of users. The profiles are provided to SSPs, AdXs and/or DSPs, which enrich the information included in ad requests and/or bid requests, allowing the delivery of personalized ads. 

Once the browser (or mobile app) receives the URL of the ad, the \emph{device} part of the ad delivery process starts (i.e., the ad rendering). The browser (or mobile app) download the ad, which is placed in the ad space, typically embedded in an iFrame (or even a double iFrame). The ad typically includes JavaScript code that is used for multiple purposes, including reporting delivery, monitoring user's actions (e.g., clicks on the ad), measuring KPIs \new{-Key Performance Indicators-} (e.g., ad viewability \cite{jicwebs:viewability, iab:viewability}) or identifying fraudulent practices \cite{pastor2019nameles, Marciel:youtube:www}.

In this paper, we focus on measuring the energy consumption of the \emph{device} part, i.e., the ad rendering process. 

\subsection{Measuring Energy Consumption of Digital Devices}
\label{subsec:energy_consumption}

Digital devices are, in essence, a set of hardware components used to execute software. The energy is consumed by hardware components, which directly depends on the demand of resources of the executed software, e.g., software requiring 100\% use of CPU leads to higher energy consumption than software using 10\% of CPU. 

There are two main methods to measure the energy consumed by a device or by specific software within a device. The first, and most accurate one, is using an energy meter tool~\cite{noureddine2013review}. These tools measure the actual energy consumed by hardware devices. Therefore, if one can configure a setup where the device runs the software to be measured exclusively, the energy consumption measured by the meter is the energy consumed by such software. However, this method is not recommended to be used for software measurements.
When software execution cannot be isolated, a common alternative is to measure the difference in the energy consumed when the device operates running and not running the software. The difference in the measured energy is considered a good approximation to the energy consumed by the software under analysis. 

However, energy meters are not always practical. They have a minimum resolution, so energy consumption below that resolution cannot be measured. For instance, if the resolution of an energy meter is 10$^{-3}$ Watts, then any software consuming less than this cannot be measured with this meter. In these cases, there is a commonly used estimation method: the number of CPU instructions executed by the software is counted~\cite{garcia2019estimation}. The chipset manufacturer datasheet provides detailed information with respect to the consumption of energy of the CPU chipset under different conditions (e.g., \% of the use of the CPU). Hence, an estimation of the energy consumed by a device in executing a given software can be obtained. Note that this approach assumes that the CPU is the hardware element consuming most of the energy in a device. The research community has developed open-source libraries that implement this method, like CodeCarbon \cite{codecarbon}. 
\section{Dataset}
\label{sec:dataset}

The first task of this paper is to define an accurate model of the energy consumption by the rendering process of an individual ad. To this end, we need to create a ground-truth dataset providing the energy consumed by real ads rendered on websites. 

We gathered two different sets of ads. The first one serves to create the model of the energy consumption of ad rendering processes. It is collected using an automated process that surfs the web to gather thousands of different advertisements. We refer to it as \emph{automated} dataset. 

To validate the generality of the model obtained from the \emph{automated} dataset, we obtain a second dataset formed by ads shown to real users while they browse the web. We refer to this dataset as \emph{human} dataset. 

The remainder of the section describes the data collection procedure and our final datasets in detail.

\subsection{Automated Dataset}
To obtain a large number of online ads, we created an automated methodology that navigates through different websites and collects the ads present on the sites. This pool of ads will be utilized later to emulate the rendering of online ads and measure their energy consumption. There are four steps to the collection procedure:
\begin{itemize}
    \item[$(i)$] First, we compiled a list of websites to look for advertisements on. We used the Alexa list of the top 500 news websites and the top 500 most visited websites worldwide by Domain Authority \cite{moz}.
    \item[$(ii)$] Second, we require these websites to start displaying ads which we then can collect, so we implemented a methodology that automatically inspects the presence of cookie consent notices and accepts them. After accepting the consent notices, ads start appearing. The automated process then scrolls down the page five times (emulating the behaviour a user may have) to identify the presence of more ads on the webpage.
    \item[$(iii)$] To retrieve the ads, our automated software uses a tool that can extract the HMTL from iFrames where ads are placed and save it to a file for offline analysis. We created a browser extension using as reference the extension eyeWnder \cite{iordanou2019beyond} for this purpose. In particular, the extension analyzes the content of an HTML page, locates the iFrames that correspond to advertisements, and extracts their HTML code. 
    \item[$(iv)$]  Having the browser extension, the navigation methodology, and the list of URLs to explore, we implemented the whole routine using the headless browser Selenium \cite{selenium} and save each of the HTML from each ad in a separate file.
\end{itemize}

We created the \emph{automated} dataset using the aforementioned procedure. It consists of a collection of 25557 ads, which serves to define the model of the energy consumption of individual ad rendering processes described in Section \ref{sec:methodology}.

\begin{figure}[t]
    \centering
    \includegraphics[width=.9\columnwidth]{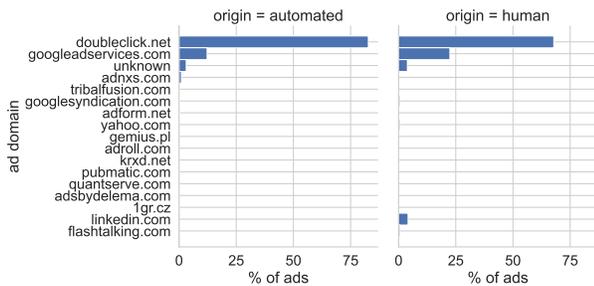}
    \caption{Distribution of ad providers from the list of 25k collected ads.}
    \label{fig:distribution_domains}
\end{figure}

\subsection{Human Dataset}

The energy consumption model described in Section \ref{sec:methodology} is obtained using our \emph{automated} dataset. These ads are obtained with automated software and thus, we have no guarantees that our model is actually representative of the energy consumption of ads shown to real users. To validate the generality of our model, we have then collected a second dataset including ads shown to real users (friends and colleagues), who have voluntarily installed a browser extension in their browser that captures the ads delivered. The \emph{human} dataset is formed by 598 ads. \Cref{fig:distribution_domains} shows the distribution of ad providers behind each ad of the \emph{automated} and \emph{human} datasets.
\section{Data-Driven Model of the Energy Consumption of Ad Rendering Process}
\label{sec:methodology}

The cornerstone piece of our system is a model, which uses parameters available upon the ad being rendered on the user's browser to estimate the energy consumption of the ad rendering process in real time. 

In this section, we thoroughly describe our methods for defining such a model. We first describe the extensive lab experiments conducted to create a ground-truth sample of ads' energy consumption using our \emph{automated} dataset, which will serve to train and validate our model. 
Afterward, we provide a detailed explanation of the process we follow to come up with our final model.

\subsection{Measuring Energy Consumption of an Ad}

The first step to build our model is to obtain a measure of the energy consumption of thousands of online ads' rendering processes, which will serve as samples of the dependent variable of our model. In the remainder of the section, we describe the procedure used to obtain an accurate estimation of the energy consumption of the ad rendering process of the 25k ads from our \emph{automated} dataset.

\subsubsection{CodeCarbon Energy Measurement Tool}

The online ad rendering process is a light software process that consumes in the order of 10$^{-6}$ KWh.\footnote{Note that this number may vary between a few 10$^{-7}$ kWh and a few 10$^{-6}$ kWh depending on the device where the ad is being rendered.} 
As discussed in Section \ref{sec:background}, commercial energy meters are unable to measure energy consumption with this granularity. Hence, following the recommendations of the literature, we opt for using a library that estimates the energy consumption based on the CPU cycles used by the software process. In particular, we use CodeCarbon \cite{codecarbon}, an open-source Python library that computes the energy and carbon footprints of any piece of code in various computer systems. CodeCarbon supports a variety of computing platforms, bases its estimations on the specific energy consumption reported by manufacturers' datasheets for tens of most commonly used CPU chipsets, and provides real time measurements. It calculates the amount of energy consumed \new{(CPU+GPU+RAM energy)} by taking into account the computer architecture, usage, and operating time. The described features make CodeCarbon an appropriate tool for achieving our goal.

Moreover, CodeCarbon is able to identify the location of a device (e.g., using the IP address) and then convert energy consumption values into carbon emission values using national grid mix data based on the country where the device is located.

\subsubsection{Lab Experiments to Measure the Energy Consumption of Ads}
\label{subsec:lab_experiments}

We first present our lab infrastructure setup and then we describe the method to isolate the ad rendering process and measure the energy consumed by it.

\paragraph{Lab infrastructure set up}
We set up a lab environment where each advertisement's energy consumption was measured using three real-world devices: a Windows desktop PC, a Linux desktop PC, and a Linux laptop. The only distinction between our two identically equipped desktop PC systems is their operating system (OS). Although the third device has different hardware than the first two, it runs the same operating system (OS) as one of them. Furthermore, every device is connected to the same network. As a result, three operating systems are available: a 64-bit Intel Windows 10 OS with 128GB RAM (Windows PC) with Google Chrome 103.0.5060.114, a 64-bit Intel Ubuntu 18.04 OS with 128GB RAM (Ubuntu PC) with Google Chrome 103.0.5060.114, and a 64-bit Intel Ubuntu 18.04 OS with 8GB RAM (Ubuntu Laptop) with Google Chrome 94.0.4606. The utilization of these different configurations aims to minimize the dependence of our final model on specific operating systems or devices.
Moreover, we obtained 2 more devices for validating our model: a Mac laptop and a Windows Laptop. Our purpose is the ML model to work on unseen devices. Therefore, we render real ads in these new devices and operating systems to validate our model. In conclusion, the validation devices consist of the three mentioned above (Windows PC, Ubuntu PC, and Ubuntu Laptop) plus an HP Laptop 64-bit Intel Core i7-1065G7 Windows 10 Home OS with 12GB RAM (Windows Laptop) and Google Chrome 106.0.5249.61; and a Intel Core i5 Macbook Pro macOS Catalina 10.15.7 with 16GB RAM (Mac Laptop) and Google Chrome 105.0.5195.125.

\begin{figure*}[!t]
    \centering
    \includegraphics[width=1.25\columnwidth]{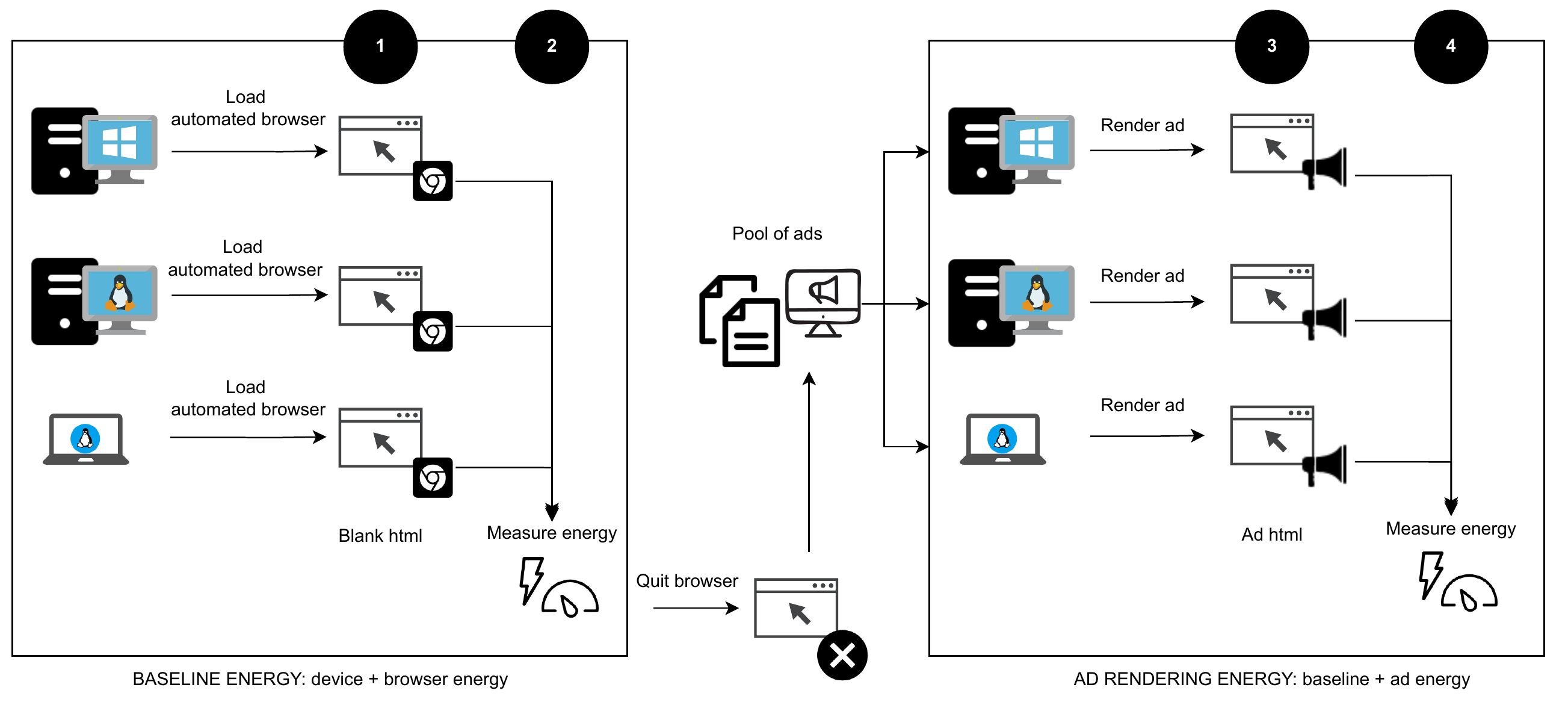}
    \caption{Diagram of the automated methodology implemented to measure the energy consumption of online ads.}
    \label{fig:measure_steps}
\end{figure*}

\paragraph{Methodology to isolate the ad rendering process}
We devised an automated method for isolating and running the rendering of each ad for each device. A diagram of the methodology is depicted in \Cref{fig:measure_steps}. In the following we describe in detail the steps taken for measuring the energy consumption of the rendering process of an individual advertisement:

\begin{itemize}
\item[$(i)$] The procedure automatically launches a new Google Chrome instance and opens a blank HTML \new{(HyperText Markup Language, standard language for web browsers)} file within it. The purpose of this step (step 1 in \Cref{fig:measure_steps}) is to obtain the baseline energy usage before loading the ad.
\item[$(ii)$] Using CodeCarbon, we obtain the energy usage of the empty instance of Google Chrome with a blank HTML. Hence, this is referred to as \emph{baseline energy} (step 2 in \Cref{fig:measure_steps}). Immediately after, we quit the browser instance and we are ready to perform again the measurement but this time loading the ad.
\item[$(iii)$] After terminating the current Google Chrome instance, the process selects one ad from the \emph{automated} dataset and launches a browser instance while rendering the advertisement. Then it waits until the ad is fully loaded.
\item[$(iv)$] Then, CodeCarbon measures the energy consumption (step 1 in \Cref{fig:measure_steps}). This measurement includes the device plus the browser consumption together with the ad's consumption. We refer to this measurement as \emph{ad rendering energy} since it measures the baseline consumption as in steps 1 and 2, together with the ad rendering process.
\end{itemize}

In summary, the described lab experiment provides two energy measurements for an individual advertisement: \emph{baseline energy}, which measures the device and browser's energy consumption, and \emph{ad rendering energy}, which measures the device, browser, and advertisements' energy consumption. We can use these two measures to define the dependent variable of our machine learning model as described in Section \ref{sec:metrics_estimator} (See Equation \ref{eq:ne}).\footnote{\new{Note that changing the order in which we measure the energy consumptions, i.e., ad rendering energy before the baseline energy, leads to the same results.}}

We repeat the process outlined above for each of the 25k ads in our \emph{automated} dataset and each of the 598 ads in our \emph{human} dataset. Furthermore, the procedure is repeated five times per ad to ensure consistent measurements, resulting in each ad measurement having five samples for \emph{baseline energy} and \emph{ad rendering energy}. Later, the median amount of energy used across these five samples is calculated. It should be noted that obtaining the energy consumption of advertisements through the use of this methodology takes time and is not immediate. On the slowest of the devices, the Windows PC, each ad measurement (i.e., the technique described above) took around 15 seconds. This means it would take us more than 4 days to collect the energy consumption measurements for the 25k advertisements. We repeated the measurements five times for accuracy and consistency, which took about 20 days of computation.

Note that during our lab experiments, for each of the 25k ads we also collect the parameters that will serve as independent variables in our ML model. Therefore, our experiments provide all the required input data for our model to estimate the energy consumption of the ad rendering process. As in the case of the energy, we collected 5 samples of the value for each parameter and consider its median value for building our model.

\subsection{Model to Estimate Energy Consumption of Ads}
\label{subsec:model}
Once we have measured the energy consumption of the 25k advertisements in our \emph{automated} dataset, we intend to develop an ad energy consumption estimator that can forecast the consumption of each ad based solely on the collected parameters during the ad rendering process. This model enables the possibility of estimating the energy consumed by each ad in real time as the advertisement is rendered, which is our main goal.

In the following, we go over the models, features, and the used ad parameters in great detail.

\subsubsection{Metric to Build the Estimator}
\label{sec:metrics_estimator}
Before continuing with the description of our model, we first introduce \emph{Ad Energy} ($E_{ad}$), a metric that directly reports the energy consumed by the rendering process of an ad. This metric is computed using the \emph{baseline energy} and \emph{ad rendering energy} described above as follows:

\begin{figure*}[t]
    \centering
    \begin{minipage}{.48\hsize}
    \centering
        \includegraphics[width=.9\columnwidth]{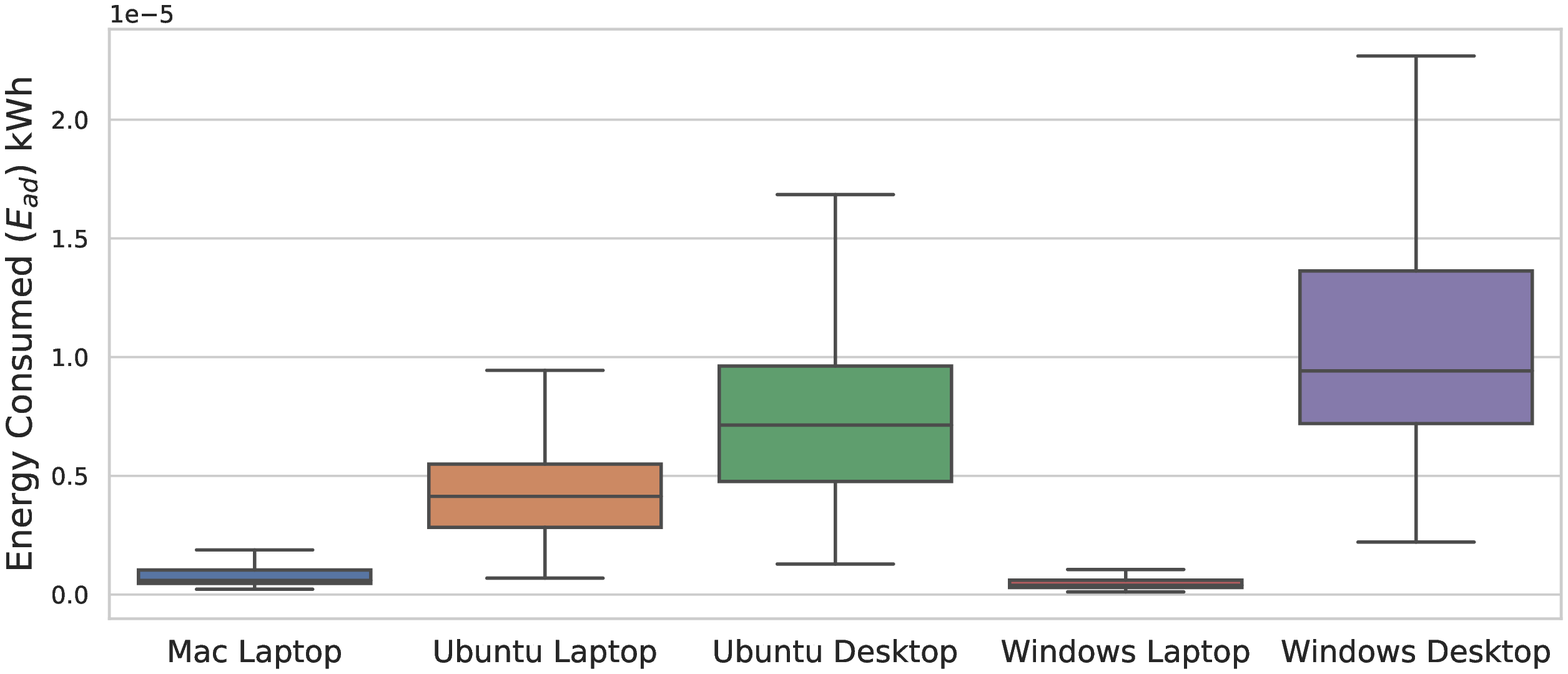}
        \caption{Energy Delta of each \emph{human} ad for the different devices analyzed.}
        \label{fig:initial_delta}
    \end{minipage}
    \hfill
    \begin{minipage}{.48\hsize}
    \centering
        \includegraphics[width=.9\columnwidth]{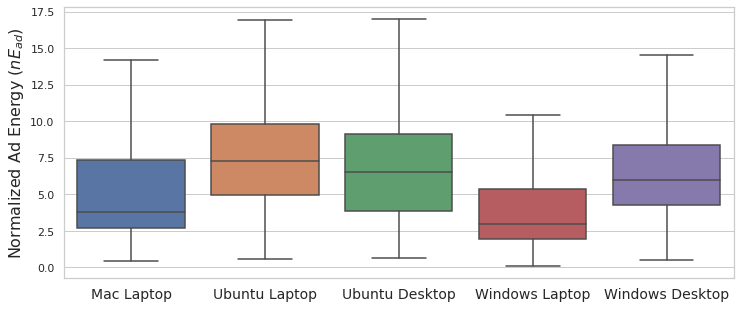}
        \caption{Normalized Ad Energy of each \emph{human} ad for the different devices analyzed.}
        \label{fig:normalized}
    \end{minipage}
\end{figure*}

\begin{equation}
    E_{ad} = ad\,rendering\,energy - baseline\,energy 
\label{eq:edelta}
\end{equation}

Using \emph{Ad Energy} ($E_{ad}$) we can report the ad energy consumption within the device on which the ad is running. This is to say, the actual energy (kWh) consumed by the advertisement rendering process on that specific device. 

\Cref{fig:initial_delta} depicts a boxplot showing the energy consumed by the rendering process of each ad from the \emph{human} dataset in each device used in our lab test environment. We can observe that the standard energy used by a certain hardware+software system differs from another hardware+software combination. The hardware in a laptop, which is designed to use less energy, is different from that in a desktop which is designed to be powered by an external source. For this reason, we determined that it is necessary to measure the relative energy consumption of an advertisement rather than the actual energy consumption ($E_{ad}$), which depends on the hardware to be measured. In order to do that, we compute the advertisement's relative energy usage, to reduce the effect of the hardware on which it runs. Therefore, if we have two ads, $A$ and $B$, we will not know exactly how much energy each of them uses (since it will vary depending on the hardware), but we will know that A uses more energy, and generally how much more energy, than B.

Hence we present \emph{Normalized Ad Energy} ($nE_{ad}$), a metric that informs about the relative use of energy of the ad rendering processes with respect to the baseline energy consumption of such device:

\begin{equation}
    nE_{ad} = \frac{ad\,rendering\,energy - baseline\,energy}{baseline\,energy}
\label{eq:ne}
\end{equation}
\emph{Normalized Ad Energy} acts as the dependent variable (i.e., the variable to be predicted) for our model. \new{\Cref{fig:normalized} depicts a boxplot showing the distribution of the normalized energy of the rendering process of each ad from the \emph{human} dataset in each device. We observe that using $nE_{ad}$ the effect of the hardware is 10 times reduced in comparison to \Cref{fig:initial_delta}, and therefore, we can achieve a good model performance with R\textsuperscript{2} values between $0,67$ and $0,97$ (See \Cref{tab:validation}).}

\subsubsection{Feature Selection}
As indicated above, our lab experiment will collect for the 25k ads from our \emph{automated} and \emph{human} datasets the list of parameters described in \Cref{tab:parameters}, which are the initial set of independent variables under consideration to be used in our model. We undergo a feature selection process, to make the final selection of this initial set of parameters that will be used in the model.

To better understand our data and to discard less relevant independent parameters, we first made a manual inspection to see the distribution of each parameter, outliers, and the correlation with the dependent variable, i.e., the normalized energy consumption. This process suggested that the following independent variables should be included: \textit{screen\_size, totalJSHeapSize,  entries, et\_paint, et\_resource, it\_xmlhttprequest,  it\_img,  it\_script, ad\_navigation\_duration, ad\_navigation\_processing, ad\_navigation\_onLoad, duration\_mean, redirectTime\_mean, request\_mean, and response\_mean}. 

It is relevant to highlight that in this manual process we have analyzed the importance of each variable. For instance, the ad size, by checking the correlation between $nE_{ad}$ and \emph{ad\_navigation\_transferSize}. This variable represents the size of the fetched ad in bytes. As a result, the \emph{automated} ads dataset is distributed by ads whose size is 28kB, 41kB, and 49kB in the 25, median, and 75 quantiles, respectively. For the \emph{human} ads dataset this values are 48kB, 58kB, 75kB, in the 25, median, and 75 quantiles, respectively. The Pearson correlation between $nE_{ad}$ and \emph{ad\_navigation\_transferSize} is 0.0002, and 0.03 for the \emph{automated} and \emph{human} datasets, respectively. We then conclude that ad transferSize is therefore not important in the ad's rendering energy consumption.

Moreover, before developing our model, we performed an automatic feature selection process on these pre-selected independent variables to prevent over-fitting and so improve the performance of our estimator.

First, we keep the independent variables showing the highest correlation with the dependent variable ($nE_{ad}$), filtering out those variables with a correlation $<$ 0.8. 

Later, we quantify the multicollinearity using Variance Inflation Factors (VIF) to avoid considering independent variables with a high correlation with one another. We eliminate the variables with a VIF $>$ 10, which indicates strong multicollinearity. \cite{vittinghoff2006regression}

Furthermore, because predicting energy consumption can benefit from the interaction of n of two or more variables, we create new independent variables to account for all potential two and three-variable interactions.
Note that these new variables undergo the described filtering process, i.e., those showing a correlation $<$ 0.8 with $E_{ad}$ or with a VIF $>$ 10 are removed. 

Finally, we eliminate constant or almost constant variables; i.e, the variables with a variance of less than 0.01 since they won't help us estimate energy consumption better.

\begin{table*}[t]
\centering
\begin{adjustbox}{width=.9\textwidth}
\begin{tabular}{r|l}
Parameter & Description \\ \hline
usedJSHeapSize & The currently active segment of JS heap, in bytes. \\
totalJSHeapSize & The total allocated heap size, in bytes. \\
entries & Number of PerformanceEntry objects in the ad \\
entries\_requested & Number of PerformanceEntry objects with a performance duration \textgreater 0 \\
screen\_size & Screen width x screen height \\
et\_element & Number of element entryType entries executed in the ad \\
et\_navigation & Number of navigation entryType entries executed in the ad \\
et\_resource & Number of resource entryType entries executed in the ad \\
et\_mark & Number of mark entryType entries executed in the ad \\
et\_measure & Number of measure entryType entries executed in the ad \\
et\_paint & Number of paint entryType entries executed in the ad \\
et\_longtask & Number of longtask entries executed in the ad \\
it\_element & Number of element initiatorType entries executed in the ad \\
it\_css & Number of css initiatorType entries executed in the ad \\
it\_embed & Number of embed initiatorType entries executed in the ad \\
it\_img & Number of img initiatorType entries executed in the ad \\
it\_link & Number of link initiatorType entries executed in the ad \\
it\_object & Number of object initiatorType entries executed in the ad \\
it\_script & Number of script initiatorType entries executed in the ad \\
it\_subdocument & Number of subdocument initiatorType entries executed in the ad \\
it\_svg & Number of svg initiatorType entries executed in the ad \\
it\_xmlhttprequest & Number of xmlhttprequest initiatorType entries executed in the ad \\
it\_navigation & Number of navigation initiatorType entries executed in the ad \\
it\_other & Number of other initiatorType entries executed in the ad \\
duration\_mean & mean duration of the performance entries in ms \\
transferSize\_mean & mean transfer size of the performance entries of the fetched resource \\
dedodedBodySize\_mean & mean transfer size of the performance entries received from the fetch (HTTP or cache) of the message body \\
redirectTime\_mean & mean duration of the redirect phase of entries in ms \\
app\_cache\_mean & mean duration of the app cache phase of entries in ms \\
dns\_mean & mean duration of the dns phase of entries in ms \\
tcp\_mean & mean duration of the tcp phase of entries in ms \\
request\_mean & mean duration of the request phase of entries in ms \\
response\_mean & mean duration of the response phase of entries in ms \\
ad\_navigation\_duration & Time in ms of the ad's execution \\
ad\_navigation\_transferSize & A number representing the size of the ad \\
ad\_navigation\_decodedBodySize & A number that is the size (in bytes) received from the fetch (HTTP or cache) of the message body \\
ad\_navigation\_app\_cache & Duration of the APP CACHE phase of the ad in ms \\
ad\_navigation\_dns & Duration of the DNS phase of the ad in ms \\
ad\_navigation\_tcp & Duration of the TCP phase of the ad in ms \\
ad\_navigation\_request & Duration of the REQUEST phase of the ad in ms \\
ad\_navigation\_response & Duration of the RESPONSE phase of the ad in ms \\
ad\_navigation\_processing & Duration of the PROCESSING phase of the ad in ms \\
ad\_navigation\_onLoad & Duration of the ON LOAD phase of the ad in ms \\ \hline
\end{tabular}
\end{adjustbox}
\vspace{.5em}
\caption{Variables collected of each ad for analysing and populating the models.}
\label{tab:parameters}
\vspace{-3mm}
\end{table*}

\begin{table}[!ht]
\centering
\begin{adjustbox}{width=.85\columnwidth}
\begin{tabular}{r|l}
variable & VIF value \\
\hline
ad\_navigation\_duration screen\_size request\_mean & 5.78 \\
ad\_navigation\_duration ad\_navigation\_onLoad & 1.17 \\
response\_mean screen\_size & 5.12 \\
ad\_navigation\_duration redirectTime\_mean & 2.94 \\
ad\_navigation\_duration & 4.32 \\
screen\_size & 0.60 \\
tcp\_mean & 1.44 \\
request\_mean & 2.94 \\
response\_mean & 8.42 \\
it\_xmlhttprequest & 1.04 \\
redirectTime\_mean & 2.77 \\ \hline
\end{tabular}
\end{adjustbox}
\vspace{.5em}
\caption{Variables used to build the ads energy estimator based on the information collected when simulating the rendering phase of each ad.}
\label{tab:variables_model}
\end{table}

\begin{figure}[!t]
    \centering
    \includegraphics[width=.9\hsize]{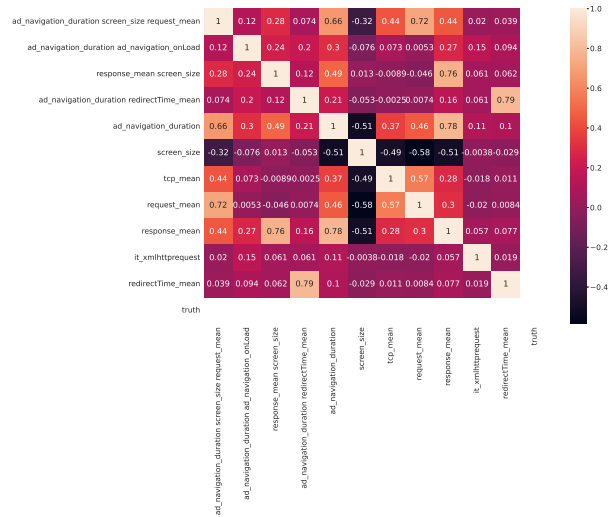}
    \caption{Correlations between each pair of independent variables used in the model.}
    \label{fig:correlations}
\end{figure}

\Cref{tab:variables_model} shows the final list of independent variables we use as input to build our energy estimator model with their respective VIF factor. \Cref{fig:correlations} depicts the correlation between each pair of variables.

\subsubsection{Model}

We want to build a model that, based solely on an ad's parameters available upon rendering the ad, can predict how much energy will be consumed by rendering it. We trained our model using the 25k ads from our \emph{automated} dataset. Our purpose is to obtain a sufficiently accurate model that can estimate the energy consumed by ads in nearly real time.

As discussed above, the \emph{automated} dataset used to build our model includes ads that are not typically shown to users (due to the fact that our method involves the collection of ads by software and not by user browsers) and thus it may present some biases and not be fully representative of ads shown to actual users. To assess the generality of our model, we validate it with the 598 ads from our \emph{human} dataset. We compute the $nE_{ad}$ of each of the ads in our \emph{automated} and \emph{human} datasets in our lab experiment setup as described in Section \ref{subsec:lab_experiments}.

\begin{table}[!t]
\centering
\begin{adjustbox}{width=.87\columnwidth}
\begin{tabular}{ccccc}
Validation & Model & Linear & Random Forest & Gradient Boosting \\
\hline
\multirow{2}{*}{Windows Desktop} & R2 & 0,97 & 0,56 & 0,56 \\
 & RMSE & 2,16 & 6,02 & 6,00 \\
 \hline
\multirow{2}{*}{Ubuntu Desktop} & R2 & 0,89 & 0,27 & 0,27 \\
 & RMSE & 6,27 & 14,44 & 14,41 \\
 \hline
\multirow{2}{*}{Ubuntu Laptop} & R2 & 0,93 & 0,88 & 0,91 \\
 & RMSE & 2,93 & 2,46 & 2,13 \\
 \hline
\multirow{2}{*}{Window Laptop} & R2 & 0,67 & 0,36 & 0,42 \\
 & RMSE & 5,78 & 6,81 & 6,48 \\
 \hline
\multirow{2}{*}{Mac Laptop} & R2 & 0,89 & 0,62 & 0,66 \\
 & RMSE & 2,92 & 3,82 & 3,66 \\ \hline
\end{tabular}
\end{adjustbox}
\vspace{.5em}
\caption{Validation with real ads.}
\vspace{-9mm}
\label{tab:validation}
\end{table}

After analyzing different models presented in \Cref{tab:validation}, we conclude that the best estimator for $nE_{ad}$ is a linear model. \Cref{tab:validation} displays the RMSE \new{(Root Mean Square Error)} and R\textsuperscript{2} scores for the different considered models\new{, these metrics are frequently used for calibrating the models, measuring the differences between the predicted and the observed values}. In this table, we present the results of comparing our model's predictions to actual readings from the 5 devices used in our validation setup. As shown in the table, our model performs sufficiently well regardless of which of the 5 devices it is compared against. Our estimator for determining the energy consumption of an online advertisement is effective, as evidenced by our Linear Regression model. One of the key advantages of our method is that we were able to develop a transversal estimator that works in a variety of relevant setups.

\subsection{Limitations}

\subsubsection{What About the Mobile Ecosystem?}
CodeCarbon, \new{the library used for measuring the real energy consumption,} does not work on mobile devices. This limitation may represent a concern. However, one of the essential goals of defining a normalized measure ($nE_{ad}$) is that it is related to the device's usage, taking into consideration the increase in energy of the ad relative to the device's base consumption. As a result, \new{although mobile ads are not used to train the model, this metric is applicable to measure the ad rendering energy consumption on mobile devices and any other hardware/OS combination not mentioned in this article.

The results obtained for the Windows Laptop and Mac Laptop, not included in the training dataset (\emph{automated} dataset), indicate that the energy consumption measurements based on our model can be accurate in other devices, including cell phones. Hence, since our pixel code works within the ad, independently of the device used, the CarbonTag estimator would get an estimation of the energy consumed by the ad rendering process on any ad, even on mobile devices. We intend to measure ad rendering energy consumption in the mobile ecosystem in future work; however, due to the complexity of doing so on a mobile device, we present a global method that can be extrapolated to any device.}

\subsubsection{Other type of ads}
\new{Our research suggests that the estimation and validation of energy consumption in the rendering process of display ads is a critical first step toward reducing the carbon footprint of online advertising. As part of our future work for CarbonTag, we plan to extend our estimates to other types of ads, such as video ads. Currently, our paper estimates and validates the energy consumed in the rendering process of display ads. However, we conjecture that CarbonTag still provides insightful information on the energy consumed by the rendering process or other types of ads, creating awareness in the community.}

\section{Ad Energy Labeling System}

Making the ad industry and the general public aware of complex processes such as the energy consumption of the ad rendering process is difficult. However, we have examples showing us the path to create this awareness. For instance, the EU's use of energy labeling framework \cite{eu_labeling} has proven that a simplified labeling system on the energy consumption of appliances creates the right awareness in the general public. Following this successful use case, we propose a labeling system for ads. This system (or an alternative one) could be used by the \adtech industry and users to filter out ads that consume an excessive amount of energy.  

\begin{figure}[t]
    \centering
    \includegraphics[width=.9\columnwidth]{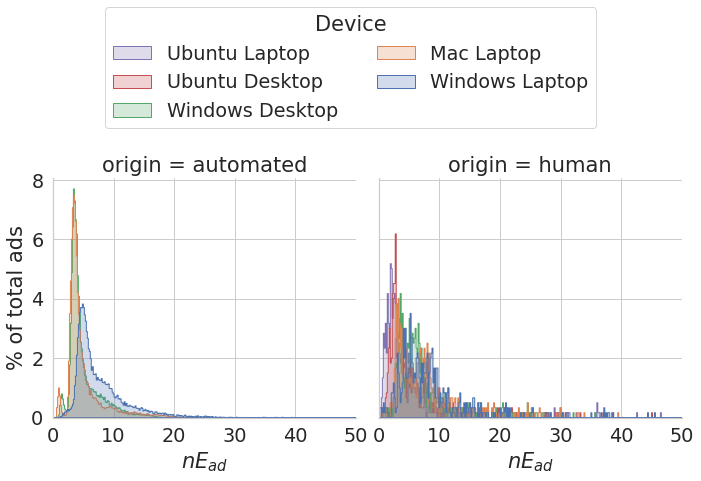}
    \caption{Distribution of energy consumption of ads by the device. On the left chart, the distribution of energy of ads gathered from automation. On the right, real ads gathered from real users.}
    \label{fig:distribution_energy_ads}
\end{figure}

We use data from the energy consumption of the 25k ads from our \emph{automated} dataset to develop a scale on which ads can be labeled as more or less energy efficient. \Cref{fig:distribution_energy_ads}depicts the breakdown of energy consumption for ads based on device type. The left chart depicts the distribution of energy usage for automated ads, whereas the right chart depicts the distribution of energy usage for real ads collected from actual users. According to it, the majority of the rendering of the advertisement is between 0-10 $nE_{ad}$.
We propose to use a simple labeling scale system based on these distributions. We mimic the EU energy labeling framework and create 7 categories, from A (most energy efficient) to G (least energy efficient). 

We propose the following labeling process according to the distributions shown in \Cref{fig:distribution_energy_ads}. $A$: 0-1 $nE_{ad}$, $B$: 1-3 $nE_{ad}$, $C$: 3-6 $nE_{ad}$, $D$: 6-10 $nE_{ad}$, $E$: 10-15 $nE_{ad}$, $F$: 15-25 $nE_{ad}$, $G$: $>$25 $nE_{ad}$.
\section{\carbonTag system}
\label{sec:system}

\carbonTag is the system that we have implemented to accomplish our goal, i.e., measuring the energy consumption of the ad rendering process in the device in real time and at scale. In this section, we describe two potential practical implementations of \carbonTag referred to as \textit{\carbonTag Server-based} and \textit{\carbonTag Serveless}. In addition, we conduct the performance evaluation to assess whether it meets the requirements imposed by the current online advertising ecosystem. To conclude the section, we discuss its readiness and flexibility to be operational. 

\subsection{\carbonTag Server-based}

\begin{figure}[t]
    \centering
    \includegraphics[width=0.7\columnwidth]{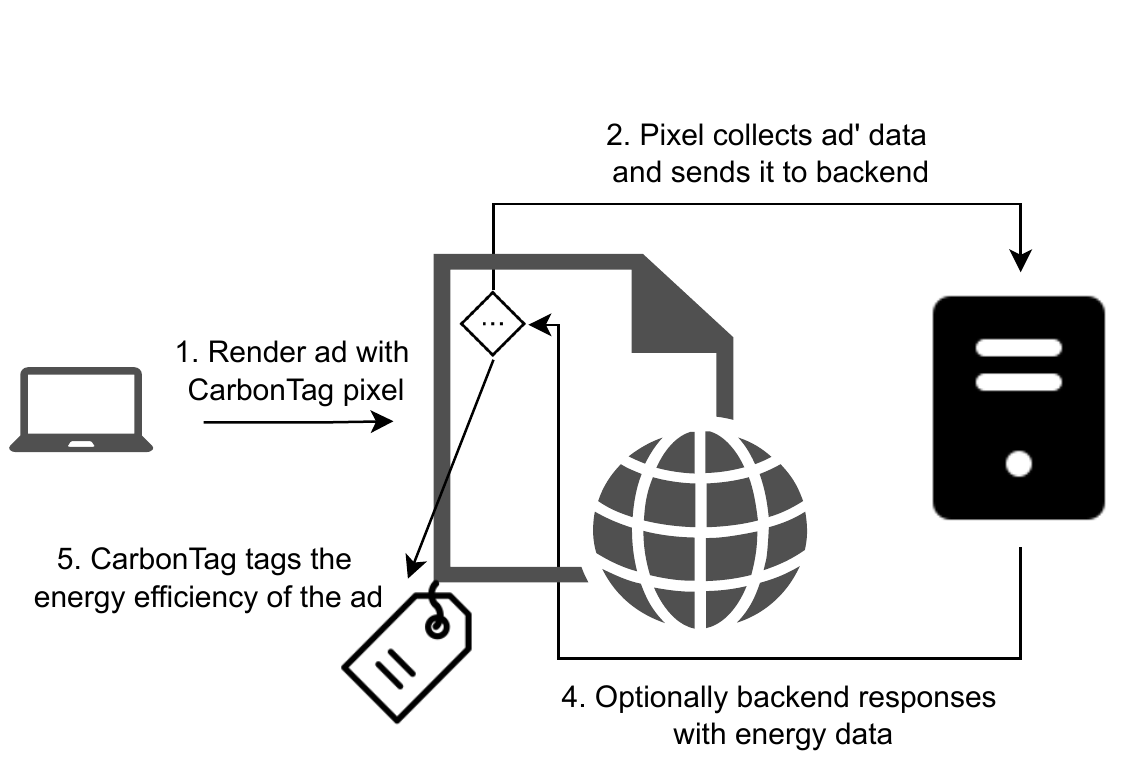}
    \caption{Visual representation of \carbonTag functionality.}
    \label{fig:diagrama_carbontag}
\end{figure}

This proposal represents the standard implementation of \carbonTag. \carbonTag Server-based is formed by two main components: the \adtag and the backend server. \Cref{fig:diagrama_carbontag} shows a visual representation of our system.

Ad tags are a standard technique used to insert code in ads or in the iFrame(s) containing the ad. These ad tags are typically JavaScript code that interacts with the browser to make different types of measurements and collect data, which are then reported to third-party servers.

In \carbonTag, the \adtag is a JavaScript code responsible for collecting the parameters (described in Section \ref{sec:methodology}) to be fed to the model as independent variables. The parameters are gathered from native JavaScript libraries, mainly based on the Performance interface~\cite{api_performance}. This library provides access to performance-related information for the ad, using multiple APIs, such as Navigation Timing API, Resource Timing API, etc., which report the information about the loading time of each resource and its associated metrics.

The \adtag sends the collected parameters to the backend server, which upon the reception of the parameters runs the model and obtains an estimation of the energy consumption and the energy label of the specific ad being measured. These results are stored locally and eventually, the server can also send the results back to the \adtag (for instance) to show the energy label to the user. 

\subsection{Serverless \carbonTag}
\label{subsec:serverless}
We have implemented an alternative version of \carbonTag, which does not use a backend server. In this version, the \adtag collects the parameters for the model and also runs the model locally in the device to obtain the estimation of the energy consumption of the ad rendering process. 

The serverless version of \carbonTag offers few advantages over the standard version. On the one hand, it requires fewer compute resources, since it does not require the implementation of any sort of communication with a third-party, and the backend infrastructure of the system is not needed. On the other hand, it limits the amount of data shared by ads with third parties, which enhances privacy.

On the downside, the serverless version requires encoding the code of the model within the \adtag~, thus not allowing to embed new heavier models that could appear in the future to estimate energy consumption. Also, the serverless version does not send ads' information to a backend, which could be useful for statistics and future analysis of ads performance or model retraining. Since the \adtag is by definition a light piece of code, the model encoded must be likewise light, limiting the type of models to be used to simple ones like linear regressors. The linear regression model can be implemented using around 10KB offering an R$^2$ between 0,67 and 0,97 (See Table \ref{tab:validation}). Instead, more complex models such as decision trees require at least a MB to be embedded in a pixel, which is not feasible.

\subsection{Performance evaluation}

The ultimate goal of \carbonTag is to measure ad energy consumption in real time and at scale. This requires meeting two specific requirements that the \adtech ecosystem would impose on our solution in order to be valid in an operational environment: real time operation and scalability. Finally, another important requirement is that \carbonTag should be very light in terms of energy consumption.

\subsubsection{Real time performance}

To evaluate the real time requirement in the server-based version, we have measured the time elapsed since the instant the ad rendering process starts until an estimation of the energy consumed by the ad is received by the \adtag. This process considers the time required by the \adtag to collect the parameters, the communication with the backend server, the time required by the backend server to run the model, and the communication of the result to the \adtag. Similarly, we have also measured the time required by the serverless solution to obtain the energy consumption metric. We have emulated the described process in our lab setup for the \emph{human} dataset. 

\begin{figure}[t]
    \centering
    \includegraphics[width=.78\columnwidth]{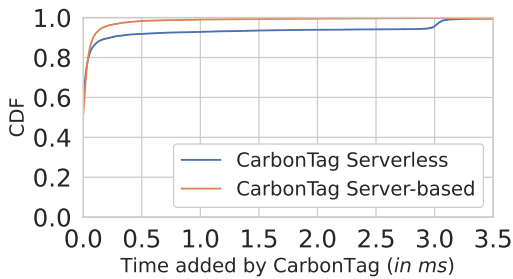}
    \caption{Cumulative Distribution Function of the time taken by \carbonTag to get the parameters and process the model, under two scenarios, online vs offline.}
    \label{fig:carbontag_timing}
\end{figure}

Figure \ref{fig:carbontag_timing} shows the CDF of the time required by \carbonTag to perform the energy consumption evaluation both in the Server-based and Severless less versions. In particular, the figure shows the difference between the load time of the ad with and without \carbonTag, which captures the time overhead added by the use of \carbonTag. We observe that the median is 0.0~ms for both \carbonTag implementations. Moreover, the 90-percentile is 0.2~ms and 0.1~ms for the server-based and  the serverless version, respectively. Our results confirm that \carbonTag can operate in real time adding a negligible timing overhead over the current operation of ads rendering process. 

\subsubsection{Scalability}

The server-based architecture of \carbonTag is a standard model used in \adtech (and the Web in general), which can easily scale with the number of requests. We have made a stress test in our lab to assess the expected number of queries that a commodity machine acting as \carbonTag backend server may handle. In particular, we used a  machine with 56~Cores, and 126~GB RAM, in a 10~Gbps connection at our lab. We clarify that the serverless solution runs locally and the requests scalability issue does not apply in this case.

We created a flow of \adtag requests at different rates. We assessed that our commodity server was able to handle \adtag requests rates generating up to 940~Mbits/sec of bandwidth. This result confirms \carbonTag offers the required scalability to operate in the current programmatic advertising ecosystem.

\subsubsection{Energy consumption}
A solution meant to measure energy consumption, must guarantee a close to negligible energy footprint. Hence, \carbonTag should present a low energy consumption.

\begin{figure}[t]
    \centering
    \includegraphics[width=.95\columnwidth]{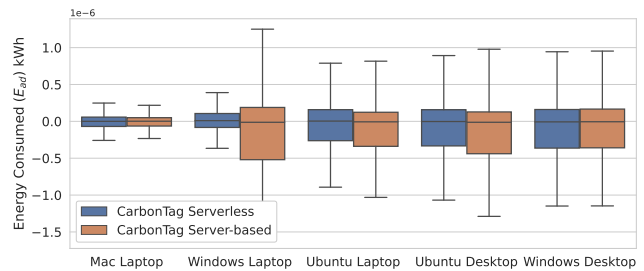}
    \caption{Performance of the solution \carbonTag, in both scenarios, loading the model online vs loading the model offline. Represents the consumption compared with the Ad. }
    \label{fig:carbontag_performance}
\end{figure}

To assess this,  we have measured the energy consumption of the ads from our \emph{human} dataset with and without \carbonTag. Figure \ref{fig:carbontag_performance} shows the distribution of the extra energy consumed by ads when running \carbonTag for the different devices considered in our lab environment both in the server-based and serverless implementations. We conclude that the impact of \carbonTag is negligible in terms of energy consumption. Indeed, for a representative set of ads, the energy consumed by the ad including \carbonTag is lower than its equivalent without \carbonTag. This indicates that there are other factors, different from \carbonTag, which influence the energy consumption of an ad at a given instant (most likely linked to the OS and browser). 

\subsection{Readiness and flexibility}

\carbonTag is implemented using standard techniques from \adtech, i.e., and \adtag communicating with a backend server. Hence, any player from the online advertising ecosystem (advertisers, DSPs, ADXs, etc.) can seamlessly integrate \carbonTag in their regular operation. In essence, they should add an extra \adtag to their operation.

In addition, \carbonTag is very flexible and adaptable. As in any machine-learning based system, the model needs to be retrained and eventually modified. If this occurred, there are two scenarios to be considered: 1) If no new parameters are used in the model, the only change would be to implement the new version of the model; 2) If the new version of the model uses a different set of parameters, then the \adtag needs to be modified to collect the new set of parameters. In this case, we need to deploy a new version of the model and a new version of the \adtag needs to be distributed across the companies using \carbonTag. As we explained earlier, any new version of the model will be valid to run in the backend server part of the server-based solution. However, depending on the complexity and size of the new model, it may not be feasible for the serverless solution. 

\section{Global Impact of Ad Energy consumption on the device}
\label{sec:discussion}

In Section \ref{sec:methodology}, we have presented, to the best of our knowledge, the largest dataset used so far to measure the energy consumption of the ad rendering process of an ad. This is the energy that the ad delivery process uses in the device. 

According to our measurements, the average energy consumed by the ad rendering process of a single ad is 5e-6 kWh in the \emph{automated} and \emph{human} datasets. More specifically, in the \emph{human} dataset, the lowest consuming device, Windows Laptop, had an average energy consumption of 5e-7 kWh per ad, and the greatest consuming device, Windows Desktop an average of 1e-5 kWh per ad.

In an exercise of illustrating the implications of these numbers, the rendering process of 5 ads will consume an energy similar to that of a TV on standby for a minute, or a 1W light for 1 minute (a night light). Even more, the rendering process of 20 ads would consume an amount of energy similar to that consumed by a 300W fridge every second.

Moreover, despite there are no scientifically proven figures, industry reports indicate that the average person is estimated to encounter between 4,000 and 10,000 ads per day~\cite{lunio,forbes,whoofey}.
Using these values as a reference we want to illustrate the online ads rendering energy consumption. We may assume that 55\% of advertising is digital \cite{55digital}, so a user may be exposed to between 2k to 5k ad impressions per day. Considering the devices reporting the lowest and greatest energy consumption of our dataset, the ads delivered to every single user would just consume between 1e-3 kWh and 50e-3 kWh per day in the rendering process.

Furthermore, sources report that more than 5 billion people (or 63 percent of the global population) use the Internet \cite{data_reportal}. If we consider this user base as a reference for our estimation, online ads would consume between 5e6 kWh and 250e6 kWh (5 million kWh to 250 million kWh) per day on user devices. As a result, online advertisements are expected to consume between 1,8e9 kWh (1,8 billion kWh) and 91e9 kWh (91 billion kWh) per year. To put these values in context, we can compare them with the overall energy consumption of different countries. If we do this exercise, we observe that the rendering of online advertisements consumes an amount of energy in the same range as countries like Luxembourg (6,5 Billion kWh), Portugal (49 billion kWh), or Belgium (84 billion kWh) \cite{eia_energy_countries}. Some may argue that the number of ads encountered on average per day is still large. However, our purpose is only illustrative. Ads could be rendered on a web page even if we do not see them. In any case, even if we consider the number of ads displayed to a user to be 10 times lower, the rendering of online advertisements would still consume a total energy compared to small countries like Rwanda (0,3 billion kWh), Malta (2,5 billion kWh), or Kenya (8,2 billion kWh).

As we have discussed above, \new{it is expected that the device part (i.e., the ad rendering process) of the energy consumed by the ad delivery process is smaller than the correspondent network part}, which involves several important factors. Under this assumption, we conjecture that the overall energy consumed by the delivery of ads might be in the order of tens to thousand of billions kWh per year.
\section{Related Work}
\label{sec:related_work}

The increase in energy consumption brought on by the development of the ICT industry over the past few decades has to be investigated. 
The ICT industry currently accounts for a sizeable portion of global energy consumption and greenhouse gas emissions, which has an impact on the environment. Numerous studies in the literature have examined this issue and highlighted the importance of reducing the amount of energy we use in favor of more environmentally friendly methods. 
Researchers concur that emissions and energy use from digital gadgets would likely rise in the near future. The ICT emissions share footprint will quadruple from the value of 2007 to 3-3.6\% by 2020 and to 14\% of the 2016 level by 2040, according to Belkhir et al.~\cite{belkhir2018assessing}. Additionally, Andrae et al.~\cite{andrae2020hypotheses} mention predictions that CO2 emissions in the digital industry will increase between 2020 and 2030. They predict that between 82 and 96\% of the rise in CO2 emissions in this decade would be attributed to computing and the digital industry.

Individual device usage is closely tied to the rise in this energy consumption.
According to Ruiz et al.~\cite{ruiz2022life}, end-user device emissions account for the majority of the carbon footprint of wireless ICT networks. In addition, Belkhir et al.~\cite{belkhir2018assessing} predicted that by 2020, smartphones would account for 11\% of all ICT use, outpacing the contributions of desktops, laptops, and displays taken individually.

This increase in end-user carbon emissions emphasizes the need for developing tools to comprehend ICT energy consumption and moving towards more energy-efficient approaches. Monañana et al.~\cite{montanana2021towards} for example, focus on providing a tool to achieve a more efficient computation at the software level by logging the energy consumption of various algorithms and then choosing the more energy-efficient one for the requirements of an application.

The use of personal devices is inextricably linked to the use of the Internet. Furthermore, as stated in \Cref{sec:introduction}, one of the primary revenue sources for the internet is the online advertising industry. As a result, measuring the energy consumed by online ads is critical. Despite the fact that the online advertising ecosystem and the presence of available ad spaces continue to grow, according to~\cite{IAB2021}, there is very little literature on the subject.

Pärssinen et al.~\cite{parssinen2018environmental} acknowledge that measuring the energy footprint on the Internet is difficult due to the complexity of the network. Hence, calculating the energy consumption of online advertisements is not a simple task. To estimate the energy usage of any online service, they present a framework in their study as a result. They later estimate the energy usage of the online advertising ecosystem using this paradigm. They discovered that between 20 and 282 TWh of energy was consumed by online advertising in 2016. The total amount of energy used by all Internet-related mediums in that year ranged from 791 to 1334 TWh.

Similar to our proposal, two works measure the cost in terms of online advertising energy consumption. Simons ~\cite{simons2010hidden}, simulating the actual usage of the web, measures the energy consumption with the measuring device Voltcraft Energy Monitor 3000. This device can measure with a resolution of 1W. We tried this approach and concluded that the results on the devices are not deterministic in the measurements, and the method is not scalable to multiple devices and environments. 

There is another related paper from Prochkova et al.~\cite{prochkova2012energy} where the authors measure energy consumption by comparing the power usage of mobile phone games with and without advertisements. Again, this paper uses a measuring device that is not comparable to our solution, which can measure each ad in real time.

Finally, Scope3 \cite{scope3} uses a qualitative approach without the utilization of any direct measurement. That is using a number of constant variables of the energy consumption given by the different stakeholders like Google, Criteo, Pubmatic, etc. Hence, the accuracy of their estimation will depend on the accuracy of the used qualitative assumptions and the accuracy of these companies, whose methodology is not disclosed. Instead, our methodology measures the parameters associated with the ad rendering process on different devices and OSs. Moreover, while Scope3 provides energy consumption estimation at an aggregate level for different ad stakeholders, \carbonTag provides energy consumption of the ad rendering process at the granularity of individual ads. We propose that advertisers and \adtech companies will benefit from implementing both kinds of methods: those similar to Scope3 (top-down), and those similar to \carbonTag (bottom-up).

Putting all of this together, this paper presents an approach that measures the energy consumed by online advertisements, which run and consume energy from end-user devices, and provides an energy estimation \adtag to classify those advertisements that are less energy efficient, with the goal of limiting their display. To the best of the authors' knowledge, this is the first work that presents a methodology and a tool for estimating the energy consumption of on ads in real time at the individual ad level.
\section{Conclusion}
\label{sec:conclusion}
We investigated the energy consumption caused by the rendering process of online ads in this study. 
To measure an ad's energy consumption, software or hardware installation is needed. However, this approach would require significant changes to the established online advertising ecosystem and user behavior.
For these reasons, we decided to develop a methodology that transparently calculates the energy required by an ad using only the information available during the ad's rendering process, while not interfering with its current operation in any way.

As a result, we investigated the variables most closely related to energy consumption and created an model that predicts the energy consumed by an online advertisement. We examined a sample of 25k advertisements and compared our findings to 598 online ads obtained from real people browsing the Internet. For this purpose, we used software to measure the energy used by advertisements across three different devices and operating systems. Then, without using any outside sources, we created our ad energy model, which calculates consumption solely based on the advertisement's internal rendering characteristics. Overall, our model achieves a great performance with an RMSE between 2,16 and 6,27 and a R\textsuperscript{2} between 0,67 and 0,97 working across different combinations of operating systems and devices.

Finally, we created \carbonTag
an \adtag that can be used during the rendering of any online ad to estimate the energy consumption of individual ads in real time and at scale. This would allow the \adtag to determine how energy efficient a given online ad is. Furthermore, we proposed a standardized meter for classified advertising on a scale of less to more effective (from A being ads that are the most energy efficient to G being the least), as well as the development of tools by intermediaries and \adtech actors that would allow users to specify the types of ads with which they wanted to be exposed to based on their energy efficiency. This is an important aspect of our research because, according to our estimations, just the rendering part of online ads may consume the same amount of energy per year as a whole country like Luxembourg (if we consider the bottom range of our estimation) or Sweden (if we consider the top range). The network part of online ads will consume importantly more energy than the device part.

To the best of the authors' knowledge, this is the first piece of work that estimates the energy consumption associated with rendering online ads.

%\section*{Acknowledgments}
%This should be a simple paragraph before the References to thank those individuals and institutions who have supported your work on this article.
 
\bibliographystyle{ieeetr}
\bibliography{references}

\begin{thebibliography}{10}

\bibitem{bp2022}
{BP}, ``{Statistical Review of World Energy 71st edition}.''
  \url{https://www.bp.com/en/global/corporate/energy-economics/statistical-review-of-world-energy.html}.
\newblock Accessed: Jul 19, 2022.

\bibitem{eia2022}
{U.S. Energy Information Administration (EIA)}, ``{International Energy Outlook
  2021}.'' \url{https://www.eia.gov/outlooks/ieo/}, 2021.
\newblock Accessed: Jul 19, 2022.

\bibitem{hubler2013eu}
{H{\"u}bler, Michael and L{\"o}schel, Andreas}, ``{The EU decarbonisation
  roadmap 2050—what way to walk?},'' {\em Energy Policy}, vol.~55,
  pp.~190--207, 2013.
\newblock \mbox{DOI}: \url{https://doi.org/10.1016/j.enpol.2012.11.054}.

\bibitem{aslan2018electricity}
{Aslan, Joshua and Mayers, Kieren and Koomey, Jonathan G and France, Chris},
  ``{Electricity intensity of internet data transmission: Untangling the
  estimates},'' {\em Journal of Industrial Ecology}, vol.~22, no.~4,
  pp.~785--798, 2018.
\newblock \mbox{DOI}: \url{https://doi.org/10.1111/jiec.12630}.

\bibitem{websitecarbon}
{Wholegrain Digital}, ``{The original Website Carbon calculator}.''
  \url{https://www.websitecarbon.com/}, 2022.
\newblock Accessed: Jul 19, 2022.

\bibitem{conversation:internet_consumption}
{Jeff Kettle}, ``{The internet consumes extraordinary amounts of energy}.''
  \url{https://theconversation.com/the-internet-consumes-extraordinary-amounts-of-energy-heres-how-we-can-make-it-more-sustainable-160639},
  2021.
\newblock Accessed: Jul 19, 2022.

\bibitem{nature:electricity}
{Nicola Jones}, ``{How to stop data centres from gobbling up the world’s
  electricity}.'' \url{https://www.nature.com/articles/d41586-018-06610-y},
  2022.
\newblock Accessed: Jul 19, 2022.

\bibitem{IAB2021}
{The Interactive Advertising Bureau (IAB)}, ``{Internet Advertising Revenue
  Report 2021}.''
  \url{https://www.iab.com/insights/internet-advertising-revenue-report-full-year-2021/},
  2022.
\newblock Accessed: Jul. 19, 2022.

\bibitem{parssinen2018environmental}
{P{\"a}rssinen, Matti and Kotila, M and Cuevas, R and Phansalkar, A and Manner,
  Jukka}, ``{Environmental impact assessment of online advertising},'' {\em
  Environmental Impact Assessment Review}, vol.~73, pp.~177--200, 2018.
\newblock \mbox{DOI}: \url{https://doi.org/10.1016/j.eiar.2018.08.004}.

\bibitem{simons2010hidden}
{Simons, RJG and Pras, Aiko}, ``{The hidden energy cost of web advertising},''
  in {\em Proceedings of the 12th Twente Student Conference on Information
  Technology}, pp.~1--8, Citeseer, 2010.

\bibitem{prochkova2012energy}
{Prochkova, Irena and Singh, Varun and Nurminen, Jukka K}, ``{Energy cost of
  advertisements in mobile games on the android platform},'' in {\em 2012 Sixth
  International Conference on Next Generation Mobile Applications, Services and
  Technologies}, pp.~147--152, IEEE, 2012.
\newblock \mbox{DOI}: \url{https://doi.org/10.1109/NGMAST.2012.32}.

\bibitem{openrtb}
{iab.TECH LAB}, ``{OpenRTB. Version 2.6- Released April 2022}.''
  \url{https://iabtechlab.com/wp-content/uploads/2022/04/OpenRTB-2-6_FINAL.pdf},
  2022.
\newblock Accessed: Jul. 19, 2022.

\bibitem{jicwebs:viewability}
{JICWEBS}, ``{Viewability Product Principles}.''
  \url{https://jicwebs.org/wp-content/uploads/2018/07/JICWEBS_Viewability_Product_Principles_July_2018.pdf},
  2018.
\newblock Accessed: Jul 19, 2022.

\bibitem{iab:viewability}
{MRC and IAB}, ``{MRC Viewable Ad Impression Measurement Guidelines}.''
  \url{https://www.iab.com/wp-content/uploads/2015/06/MRC-Viewable-Ad-Impression-Measurement-Guideline.pdf},
  2014.
\newblock Accessed: Jul 19, 2022.

\bibitem{pastor2019nameles}
{Pastor, Antonio and P{\"a}rssinen, Matti and Callejo, Patricia and Vallina,
  Pelayo and Cuevas, Rub{\'e}n and Cuevas, {\'A}ngel and Kotila, Mikko and
  Azcorra, Arturo}, ``{Nameles: An intelligent system for real-time filtering
  of invalid ad traffic},'' in {\em The World Wide Web Conference},
  pp.~1454--1464, 2019.
\newblock \mbox{DOI}: \url{https://doi.org/10.1145/3308558.3313601}.

\bibitem{Marciel:youtube:www}
{Marciel, Miriam and Cuevas, Rub{\'e}n and Banchs, Albert and Gonz\'{a}lez,
  Roberto and Traverso, Stefano and Ahmed, Mohamed and Azcorra, Arturo},
  ``{Understanding the Detection of View Fraud in Video Content Portals},'' in
  {\em Proceedings of the 25th International Conference on World Wide Web},
  2016.
\newblock \mbox{DOI}: \url{https://doi.org/10.1145/2872427.2882980}.

\bibitem{noureddine2013review}
{Noureddine, Adel and Rouvoy, Romain and Seinturier, Lionel}, ``{A review of
  energy measurement approaches},'' {\em ACM SIGOPS Operating Systems Review},
  vol.~47, no.~3, pp.~42--49, 2013.
\newblock \mbox{DOI}: \url{https://doi.org/10.1145/2553070.2553077}.

\bibitem{garcia2019estimation}
{Garc{\'\i}a-Mart{\'\i}n, Eva and Rodrigues, Crefeda Faviola and Riley, Graham
  and Grahn, H{\aa}kan}, ``{Estimation of energy consumption in machine
  learning},'' {\em Journal of Parallel and Distributed Computing}, vol.~134,
  pp.~75--88, 2019.
\newblock \mbox{DOI}: \url{https://doi.org/10.1016/j.jpdc.2019.07.007}.

\bibitem{codecarbon}
``{Codecarbon.io}.'' \url{https://codecarbon.io/}, 2022.
\newblock Accessed: Jul 19, 2022.

\bibitem{moz}
{MOZ Inc.}, ``{Top 500 most popular websites in the world based on Domain
  Authority}.'' \url{https://moz.com/top500}, 2022.
\newblock Accessed: Jul 19, 2022.

\bibitem{iordanou2019beyond}
{Iordanou, Costas and Kourtellis, Nicolas and Carrascosa, Juan Miguel and
  Soriente, Claudio and Cuevas, Ruben and Laoutaris, Nikolaos}, ``{Beyond
  content analysis: Detecting targeted ads via distributed counting},'' in {\em
  Proceedings of the 15th International Conference on Emerging Networking
  Experiments And Technologies}, pp.~110--122, 2019.
\newblock \mbox{DOI}: \url{https://doi.org/10.1145/3359989.3365428}.

\bibitem{selenium}
{Selenium}, ``{Selenium automates browsers}.'' \url{https://www.selenium.dev/},
  2022.
\newblock Accessed: Jul 19, 2022.

\bibitem{vittinghoff2006regression}
{Vittinghoff, Eric and Glidden, David V and Shiboski, Stephen C and McCulloch,
  Charles E}, ``{Regression methods in biostatistics: linear, logistic,
  survival, and repeated measures models},'' {\em Statistics for Biology and
  Health}, 2012.
\newblock \mbox{DOI}: \url{https://doi.org/10.1007/978-1-4614-1353-0}.

\bibitem{eu_labeling}
{European Union}, ``{About the energy label and ecodesign}.''
  \url{https://ec.europa.eu/info/energy-climate-change-environment/standards-tools-and-labels/products-labelling-rules-and-requirements/energy-label-and-ecodesign/about_en},
  2022.
\newblock Accessed: Sep 20, 2022.

\bibitem{api_performance}
{Developer Mozilla. MDN}, ``{Performance}.''
  \url{https://developer.mozilla.org/en-US/docs/Web/API/Performance}, 2023.
\newblock Accessed: Apr 10, 2023.

\bibitem{lunio}
{Sam Carr}, ``{How Many Ads Do We See A Day In 2022?}.''
  \url{https://lunio.ai/blog/strategy/how-many-ads-do-we-see-a-day/}, 2021.
\newblock Accessed: Jul 19, 2022.

\bibitem{forbes}
{Jon Simpson}, ``{Finding Brand Success In The Digital World}.''
  \url{https://www.forbes.com/sites/forbesagencycouncil/2017/08/25/finding-brand-success-in-the-digital-world/},
  2017.
\newblock Accessed: Jul 19, 2022.

\bibitem{whoofey}
{Whoofey}, ``{How many ads do we see a day in 2021}.''
  \url{https://whoofey.com/blog/how-many-ads-do-we-see-a-day-2021-ad-exposure/},
  2021.
\newblock Accessed: Jul 19, 2022.

\bibitem{55digital}
{WordStream}, ``{165 Strategy-Changing Digital Marketing Statistics for
  2022}.''
  \url{https://www.wordstream.com/blog/ws/2022/04/19/digital-marketing-statistics},
  2022.
\newblock Accessed: Jul 19, 2022.

\bibitem{data_reportal}
{Data Reportal and We Are Social and Hootsuite}, ``{DIGITAL 2022: GLOBAL
  OVERVIEW REPORT}.''
  \url{https://datareportal.com/reports/digital-2022-global-overview-report},
  2022.
\newblock Accessed: Jul 19, 2022.

\bibitem{eia_energy_countries}
{U.S. Energy Information Administration (EIA)}, ``{International Electricity
  Consumption}.''
  \url{https://www.eia.gov/international/data/world/electricity/electricity-consumption},
  2022.
\newblock Accessed: Jul 19, 2022.

\bibitem{belkhir2018assessing}
{Belkhir, Lotfi and Elmeligi, Ahmed}, ``{Assessing ICT global emissions
  footprint: Trends to 2040 \& recommendations},'' {\em Journal of cleaner
  production}, vol.~177, pp.~448--463, 2018.
\newblock \mbox{DOI}: \url{https://doi.org/10.1016/j.jclepro.2017.12.239}.

\bibitem{andrae2020hypotheses}
{Andrae, Anders SG}, ``{Hypotheses for primary energy use, electricity use and
  CO2 emissions of global computing and its shares of the total between 2020
  and 2030},'' {\em WSEAS Transactions on Power Systems}, vol.~15, no.~50-59,
  p.~4, 2020.
\newblock \mbox{DOI}: \url{https://doi.org/10.37394/232016.2020.15.6}.

\bibitem{ruiz2022life}
{Ruiz, D and San Miguel, G and Rojo, J and Teri{\'u}s-Padr{\'o}n, JG and Gaeta,
  E and Arredondo, MT and Hern{\'a}ndez, JF and P{\'e}rez, J}, ``{Life cycle
  inventory and carbon footprint assessment of wireless ICT networks for six
  demographic areas},'' {\em Resources, Conservation and Recycling}, vol.~176,
  p.~105951, 2022.
\newblock \mbox{DOI}: \url{https://doi.org/10.1016/j.resconrec.2021.105951}.

\bibitem{montanana2021towards}
{Monta{\~n}ana Aliaga, Jos{\'e} Miguel and Cheptsov, Alexey and Herv{\'a}s,
  Antonio}, ``{Towards energy efficient computing based on the estimation of
  energy consumption},'' in {\em Sustained Simulation Performance 2019 and
  2020}, pp.~21--33, Springer, 2021.
\newblock \mbox{DOI}: \url{https://doi.org/10.1007/978-3-030-68049-7_2}.

\bibitem{scope3}
{Scope3}, ``{Decarbonizing Media and Advertising}.''
  \url{https://www.scope3.com/}, 2022.
\newblock Accessed: Jul 19, 2022.

\end{thebibliography}

\vfill

\end{document}